%Paper: hep-th/9410199
%From: flad@qcd.th.u-psud.fr (flad)
%Date: Wed, 26 Oct 1994 18:24:13 +0100

\magnification=\magstep1
\parskip 4pt plus2pt minus2pt

\def\bbbc{{\mathchoice {\setbox0=\hbox{$\displaystyle\rm C$}\hbox{\hbox
to0pt{\kern0.4\wd0\vrule height0.9\ht0\hss}\box0}}
{\setbox0=\hbox{$\textstyle\rm C$}\hbox{\hbox
to0pt{\kern0.4\wd0\vrule height0.9\ht0\hss}\box0}}
{\setbox0=\hbox{$\scriptstyle\rm C$}\hbox{\hbox
to0pt{\kern0.4\wd0\vrule height0.9\ht0\hss}\box0}}
{\setbox0=\hbox{$\scriptscriptstyle\rm C$}\hbox{\hbox
to0pt{\kern0.4\wd0\vrule height0.9\ht0\hss}\box0}}}}

\vglue 1.5cm

\centerline {\bf Linear Connections on the Quantum Plane}
\vskip 1.5cm

\centerline {\bf M. Dubois-Violette, \ J. Madore, \ T. Masson}
\medskip
\centerline {\it Laboratoire de Physique Th\'eorique et Hautes
Energies\footnote{*}{\it Laboratoire associ\'e au CNRS.}}
\centerline {\it Universit\'e de Paris-Sud, B\^at. 211,  \ F-91405 ORSAY}
\vskip 1cm

\centerline {\bf J. Mourad}
\medskip
\centerline {\it  Laboratoire de Mod\`eles de Physique Mathematique}
\centerline {\it Parc de Grandmont, Universit\'e de Tours, \ F-37200 TOURS}

\vskip 2cm
\noindent
{\bf Abstract:} \ A general definition has been proposed recently of a
linear connection and a metric in noncommutative geometry. It is shown
that to within normalization there is a unique linear connection on the
quantum plane and there is no metric.

\vfill
\noindent
LPTHE Orsay 94/94
\medskip
\noindent
October, 1994
\bigskip
\eject

\beginsection 1 Linear connections

There have been several models proposed of noncommutative geometries
(Connes 1986, Dubois-Violette 1988), some of which are based on quantum
groups (Woronowicz 1987, Pusz \& Woronowicz 1989, Wess \& Zumino 1990,
Maltsiniotis 1993). A definition of a linear connection which uses only
the left-module structure of the differential forms has been proposed by
Chamseddine {\it et al} (1993). An algebra of differential forms has
however a natural structure of a bimodule.  Recently linear connections
have been considered in the particular case of differential calculi
based on derivations (Dubois-Violette \& Michor 1994a,b). and more
generally (Mourad 1994) which make essential use of this bimodule
structure. We shall here apply the general definition to the particular
case of the quantum plane. We shall show that to within normalization
there is a unique linear connection on the quantum plane. The connection
is not metric compatible. There is in fact no metric in the sense we
have defined it.

We first recall the definition of a linear connection in commutative
geometry, in a form (Koszul 1960) which allows for a noncommutative
generalization.  Let $V$ be a differential manifold and let
$(\Omega^*(V), d)$ be the ordinary differential calculus on $V$. Let $H$
be a vector bundle over $V$ associated to some principle bundle $P$. Let
${\cal C}(V)$ be the algebra of smooth functions on $V$ and ${\cal H}$
the left ${\cal C}(V)$-module of smooth sections of $H$.

A connection on $P$ is equivalent to a covariant derivative on $H$.
This in turn can be characterized as a linear map
$$
{\cal H} \buildrel D \over \longrightarrow
\Omega^1(V) \otimes_{{\cal C}(V)} {\cal H}                       \eqno(1.1)
$$
which satisfies the condition
$$
D (f \psi) =  df \otimes \psi + f D\psi                          \eqno(1.2)
$$
for arbitrary $f \in {\cal C}(V)$ and $\psi \in {\cal H}$. There is an
immediate extension of $D$ to a map
$$
\Omega^*(V) \otimes_{{\cal C}(V)} {\cal H} \rightarrow
\Omega^*(V) \otimes_{{\cal C}(V)} {\cal H}
$$
by requiring that it be an antiderivation of degree 1. From (1.2) it
follows that the difference between two covariant derivatives is an algebra
morphism of ${\cal H}$.

The definition of a connection as a covariant derivative has an
immediate extension (Connes 1986) to noncommutative geometry. Let
${\cal A}$ be an arbitrary algebra and $(\Omega^*({\cal A}),d)$ be a
differential calculus over ${\cal A}$. One defines a covariant
derivative on a left ${\cal A}$-module ${\cal H}$ as a map
$$
{\cal H} \buildrel D \over \longrightarrow
\Omega^1({\cal A}) \otimes {\cal H}                               \eqno(1.3)
$$
which satisfies the condition (1.2) but with $f \in {\cal A}$. There is
again an extension of $D$ to a map
$$
\Omega^*({\cal A}) \otimes_{\cal A} {\cal H} \rightarrow
\Omega^*({\cal A}) \otimes_{\cal A} {\cal H}
$$
by requiring that it be an antiderivation of degree 1.

A linear connection on $V$ can be defined as a connection on the
cotangent bundle to $V$. It can be characterized as a linear map
$$
\Omega^1(V) \buildrel D \over \longrightarrow
\Omega^1(V) \otimes_{{\cal C}(V)} \Omega^1(V)                     \eqno(1.4)
$$
which satisfies the condition
$$
D (f \xi) =  df \otimes \xi + f D\xi                              \eqno(1.5)
$$
for arbitrary $f \in {\cal C}(V)$ and $\xi \in \Omega^1(V)$.

If, for simplicity, we suppose $V$ to be parallelizable we can choose a
globally defined moving frame $\theta^\alpha$ on $V$. The connection
form $\omega^\alpha{}_\beta$ is defined then in terms of the covariant
derivative of the moving frame:
$$
D\theta^\alpha = -\omega^\alpha{}_\beta \otimes \theta^\beta.     \eqno(1.6)
$$
Because of (1.2) the covariant derivative $D\xi$ of an arbitrary element
$\xi = \xi_\alpha \theta^\alpha \in \Omega^1(V)$ can be written as
$D\xi = (D\xi_\alpha) \otimes \theta^\alpha$ where
$$
D\xi_\alpha = d\xi_\alpha - \omega^\beta{}_\alpha \xi_\beta.      \eqno(1.7)
$$

Let $\pi$ be the projection of
$\Omega^1(V) \otimes_{{\cal C}(V)} \Omega^1(V)$ onto $\Omega^2(V)$
defined by the wedge product on the forms. The torsion form
$\Theta^\alpha$ can be defined as
$$
\Theta^\alpha = (d - \pi D)\theta^\alpha.                         \eqno(1.8)
$$

The derivative $D_X\xi$ along the vector field $X$,
$$
D_X \xi = i_X D \xi,                                              \eqno(1.9)
$$
is a linear map of $\Omega^1(V)$ into itself. In particular
$D_X \theta^\alpha =  -\omega^\alpha{}_\beta (X)\theta^\beta$. Using
$D_X$ an extension of $D$ can be constructed  to the tensor product
$\Omega^1(V) \otimes_{{\cal C}(V)} \Omega^1(V)$. We define
$$
D_X(\theta^\alpha \otimes \theta^\beta) =
D_X\theta^\alpha\otimes \theta^\beta +
\theta^\alpha \otimes D_X\theta^\beta                             \eqno(1.10)
$$
Now let $\sigma$ be the action on
$\Omega^1(V) \otimes_{{\cal C}(V)} \Omega^1(V)$ defined by the
permutation of two derivations:
$$
\sigma (\xi \otimes \eta) (X,Y) = \xi \otimes \eta (Y,X)          \eqno(1.11)
$$
and define $\sigma_{12} = \sigma \otimes 1$ as acting on the tensor
product over ${{\cal C}(V)}$ of three factors of $\Omega^1(V)$.  Then
(1.10) can be rewritten without explicitly using the vector field as
$$
D(\theta^\alpha \otimes \theta^\beta) =
D\theta^\alpha\otimes \theta^\beta +
\sigma_{12} (\theta^\alpha \otimes D\theta^\beta).                \eqno(1.12)
$$
Define $\pi_{12} = \pi \otimes 1$. If the torsion vanishes one finds that
$$
\pi_{12} D^2 \theta^\alpha = - \Omega^\alpha{}_\beta \otimes \theta^\beta
                                                                  \eqno(1.13)
$$
where $\Omega^\alpha{}_\beta$ is the curvature 2-form. Notice that
the equality
$$
\pi_{12} D^2 (f \theta^\alpha) = f \pi_{12} D^2 \theta^\alpha     \eqno(1.14)
$$
is a consequence of the identity
$$
\pi (\sigma + 1) = 0                                              \eqno(1.15)
$$

The module $\Omega^1(V)$ has a natural structure as a right
${\cal C}(V)$-module and the corresponding condition equivalent to (1.5)
is determined using the fact that ${\cal C}(V)$ is a commutative algebra:
$$
D (\xi f) =  D (f \xi).                                           \eqno(1.16)
$$
Using $\sigma$ this can also be written in the form
$$
D(\xi f) = \sigma (\xi \otimes df) + (D\xi) f.                    \eqno(1.17)
$$

By extension, a linear connection over a general noncommutative algebra
${\cal A}$ with a differential calculus $(\Omega^*({\cal A}),d)$ can be
defined as a linear map
$$
\Omega^1 \buildrel D \over \longrightarrow
\Omega^1 \otimes_{{\cal A}} \Omega^1                              \eqno(1.18)
$$
which satisfies the condition (1.5) for arbitrary $f \in {\cal A}$ and
$\xi \in \Omega^1$. We have here set $\Omega^1({\cal A}) = \Omega^1$.
The module $\Omega^1$ has again a natural structure as a right
${\cal A}$-module but in the noncommutative case it is impossible in
general to consistently impose the condition (1.16) and a substitute
must be found.  We must decide how it is appropriate to define
$D(\xi f)$ in terms of $D (\xi)$ and $df$. We propose (Mourad 1994) to
postulate the existence of a map
$$
\Omega^1 \otimes_{\cal A} \Omega^1 \buildrel \sigma \over \longrightarrow
\Omega^1 \otimes_{\cal A} \Omega^1                               \eqno(1.19)
$$
which satisfies (1.15) to replace the one defined by (1.11). We define
then $D(\xi f)$ by the equation (1.17) but using (1.19) instead of
(1.11). The extension of $D$ to $\Omega^1 \otimes \Omega^1$ is given by
(1.12) but again using (1.19).  In order that the two different ways of
calculating $D(\xi fg)$ yield the same result we must impose that
$\sigma$ be right ${\cal A}$-linear. In general
$$
\sigma^2 \neq 1.                                               \eqno(1.20)
$$

A covariant derivative is a map of the form (1.18) which satisfies the
Leibniz rules (1.5) and (1.17). Because of (1.15) the image of the operator
$d - \pi D$ acting on $\Omega^1({\cal A})$ is a bi-submodule $\Theta$
of $\Omega^2({\cal A})$. It is the torsion submodule.

The extension of $D$ to $\Omega^1 \otimes_{\cal A} \Omega^1$ is given by
the analogue of (1.12):
$$
D(\xi \otimes \eta) =
D\xi \otimes \eta + \sigma_{12}(\xi \otimes D\eta).           \eqno(1.21)
$$
We can then define the map
$$
\Omega^1 \buildrel \pi_{12} D^2 \over \longrightarrow
\Omega^2 \otimes_{\cal A} \Omega^1                            \eqno(1.22)
$$
which can be extended by a projection
$$
\Omega^2 \otimes_{\cal A} \Omega^1 \rightarrow
(\Omega^2/\Theta) \otimes_{\cal A} \Omega^1.                   \eqno(1.23)
$$
After the projection, $\pi_{12} D^2$ is left ${\cal A}$-linear:
$$
\pi_{12} D^2 (f \xi) = f \pi_{12} D^2 \xi.                     \eqno(1.24)
$$
It will not in general be right ${\cal A}$-linear. However in the
particular case which we shall consider in the next section the
right-module structure is completely determined by the left-module
structure. There is a representation $\rho$ of the algebra such that
$$
\pi_{12} D^2 (\xi f) = (\pi_{12} D^2 \xi) \rho(f)             \eqno(1.25)
$$
after the projection (1.23).

As a simple example we mention the universal calculus which has the
property that $\pi = 1$. Therefore from (1.15) we see that
$\sigma = -1$.  If the torsion is to vanish the only possible covariant
derivative then is the ordinary exterior derivative. Every torsion-free
linear connection has vanishing curvature.

A metric $g$ on $V$ can be defined as a ${\cal C}(V)$-linear, symmetric
map of $\Omega^1(V) \otimes_{{\cal C}(V)} \Omega^1(V)$ into ${\cal C}(V)$.
This definition makes sense if one replaces ${\cal C}(V)$ by an algebra
${\cal A}$ and $\Omega^1(V)$ by $\Omega^1({\cal A})$. By analogy with
the commutative case we shall say that the covariant derivative (1.17)
is metric if the following diagram is commutative:
$$
\matrix{
\Omega^1 \otimes_{\cal A} \Omega^1
&\buildrel D \over \longrightarrow
&\Omega^1 \otimes_{\cal A} \Omega^1 \otimes_{\cal A} \Omega^1         \cr
g \downarrow \phantom{g}&&\phantom{1 \otimes g}\downarrow 1 \otimes g  \cr
{\cal A} &\buildrel d \over \longrightarrow &\Omega^1
}                                                              \eqno(1.26)
$$
In general symmetry must be defined with respect to the map $\sigma$. We
impose then on $g$ the condition
$$
g \sigma = g.                                                  \eqno(1.27)
$$

\beginsection 2 The quantum plane

In this section we apply our general prescription to the quantum plane
(Manin 1989), which possesses a natural map $\sigma$. The algebra of forms
$\Omega^* = \Omega^0 \oplus \Omega^1 \oplus \Omega^2$ has 4 generators
$x^i = (x,y)$ and $\xi^i = dx^i = (\xi, \eta)$ which satisfy the relations
$$
\matrix{
xy = qyx,\hfill          &\xi^2 = 0,\hfill        &\eta^2 = 0,\hfill
&\eta \xi + q \xi \eta = 0,\hfill\cr
x\xi = q^2 \xi x,\hfill  &x \eta = q \eta x + (q^2-1)\xi y,\hfill
&y\xi = q \xi y,\hfill   &y \eta = q^2 \eta y.\hfill
}                                                               \eqno(2.1)
$$
These conditions that can be written in the form
$$
x^ix^j - q^{-1} \hat R^{ij}{}_{kl} x^kx^l = 0,                 \eqno(2.2a)
$$
$$
x^i \xi^j - q \hat R^{ij}{}_{kl}  \xi^k x^l = 0,               \eqno(2.2b)
$$
$$
\xi^i \xi^j + q \hat R^{ij}{}_{kl}  \xi^k \xi^l = 0.           \eqno(2.2c)
$$
By grouping the indices the 4-index tensor $\hat R^{ij}{}_{kl}$ can
be written as a $4 \times 4$ matrix:
$$
\hat R = \pmatrix{q  &0         &0  &0           \cr
                  0  &q-q^{-1}  &1  &0           \cr
                  0  &1         &0  &0           \cr
                  0  &0         &0  &q}.                         \eqno(2.3)
$$

If the generators $(a,b,c,d)$ of $SL_q(2,\bbbc)$ are written in the form
of a matrix
$$
a^i_j = \pmatrix{a &b \cr c &d}                                   \eqno(2.4)
$$
then the invariance of $\Omega^*$ under the action of $SL_q(2,\bbbc)$
follows from the identity
$$
\hat R^{ij}{}_{kl} a^k_m a^l_n = a^i_k a^j_l\hat R^{kl}{}_{mn}.
                                                                  \eqno(2.5)
$$
Introduce the trivial differential calculus on $SL_q(2,\bbbc)$ with the
differential $d$ given by
$$
da^i_j = 0.                                                       \eqno(2.6)
$$
The result of the coaction of $SL_q(2,\bbbc)$ on $x^i$ and $\xi^i$ is then
$$
x^{i\prime} = a^i_j \otimes x^j, \quad
\xi^{i\prime} = a^i_j \otimes \xi^j.                              \eqno(2.7)
$$
and from (2.5) it follows that $x^{i\prime}$ and $\xi^{i\prime}$ satisfy
the same relations as $x^i$ and $\xi^i$.

We introduce the 1-form
$$
\theta = x \eta - q y \xi.                                        \eqno(2.8)
$$
It is easily seen that
$$
\theta^2 = 0                                                      \eqno(2.9)
$$
and that $\theta$ is invariant under the coaction of $SL_q(2,\bbbc)$:
$$
\theta^\prime = 1 \otimes \theta.                                 \eqno(2.10)
$$
It is in fact, to within multiplication by a complex number, the only
invariant element of $\Omega^1$. From (2.1) we deduce the commutation
relations
$$
x^i \theta = q \theta x^i, \qquad \xi^i \theta = - q^{-3} \theta \xi^i.
                                                                  \eqno(2.11)
$$

To fix the definition of a covariant derivative we must first introduce
the operator $\sigma$ of Equation~(1.19). If we take the covariant
derivative of both sides of equation (2.2b) we find that $\sigma$ is
determined on $\xi^i \otimes \xi^j$. It is the inverse of the matrix
$q \hat R$. Written out in detail it becomes
$$
\matrix{
\sigma(\xi \otimes \xi) = q^{-2} \xi \otimes \xi,\hfill      &
\sigma(\xi \otimes \eta) = q^{-1} \eta \otimes \xi,\hfill    \cr
\sigma(\eta \otimes \xi) =
q^{-1} \xi \otimes \eta - (1-q^{-2}) \eta \otimes \xi,\hfill &
\sigma(\eta \otimes \eta) = q^{-2} \eta \otimes \eta.\hfill
}                                                                \eqno(2.12)
$$
The extension to $\Omega^1 \otimes_{\Omega^0} \Omega^1$ is given by the
right $\Omega^0$-linearity. In fact, in this case $\sigma$ is also left
$\Omega^0$-linear. One verifies immediately that the condition (1.15) is
satisfied. As a result of the linearity one finds
$$
\matrix{
\sigma(\xi \otimes \theta) = q^{-3} \theta \otimes \xi,\hfill      &
\sigma(\theta \otimes \xi) = q \xi \otimes \theta
                           - (1-q^{-2}) \theta \otimes \xi,\hfill \cr
\sigma(\eta \otimes \theta) = q^{-3} \theta \otimes \eta, \hfill   &
\sigma(\theta \otimes \eta) = q \eta \otimes \theta
                            - (1-q^{-2}) \theta \otimes \eta,\hfill
}
$$
as well as
$$
\sigma(\theta \otimes \theta) = q^{-2} \theta \otimes \theta.
$$

Although $\sigma^2 \neq 1$, one finds that $\sigma$ satisfies the Hecke
relation
$$
(\sigma + 1)(\sigma - q^{-2}) = 0.
$$
The eigenvectors are
$\xi \otimes \xi$, $\eta \otimes \eta$ and
$\xi \otimes \eta + q^{-1} \eta \otimes \xi$
corresponding to $q^{-2}$ and $\xi \otimes \eta - q \eta \otimes \xi$
corresponding to $-1$. The exterior algebra (symmetric algebra) is
obtained by dividing the tensor algebra by the ideal generated by the
eigenvectors of $q^{-2}$ (the eigenvector of $-1$). The braid relation
$$
\sigma_{12}\sigma_{23}\sigma_{12} = \sigma_{23}\sigma_{12}\sigma_{23}
$$
assures us that on the tensor product of three copies of $\Omega^1$ the
two ways of taking the product yield the same answer. We shall however
not explicitly use this fact.

There is a unique one-parameter family of covariant derivatives
compatible with the algebraic structure (2.1) of $\Omega^*$. It
is given by
$$
D\xi^i = \mu^4 x^i \theta \otimes \theta.                       \eqno(2.13)
$$
The parameter $\mu$ must have the dimensions of inverse length. From the
invariance of $\theta$ it follows that $D$ is invariant under the
coaction of $SL_q(2,\bbbc)$.  From Equation~(2.9) one sees that the
torsion vanishes.

Using (1.21) one finds the equality
$$
\pi_{12} D^2 \xi^i =  \Omega^i \otimes \theta                   \eqno(2.14)
$$
where the 2-form $\Omega^i$ is given by
$$
\Omega^i =  \mu^4 q^{-2}(q^2 + 1)(q^4 + 1) x^i \xi \eta.        \eqno(2.15)
$$
It vanishes for $q = \pm i$ and $q^2 = \pm i$ but it does not vanish
when $q = 1$. There is a preferred family of non-trivial linear
connections on the ordinary complex 2-plane which are stable under the
quantum deformation. Equation (2.14) can also be written in the usual
form analogous to (1.13)
$$
\pi_{12} D^2 \xi^i = - \Omega^i{}_j \otimes \xi^j,               \eqno(2.16)
$$
with the curvature 2-form given by
$$
\Omega^i{}_j = \mu^4 (1+ q^{-2})(1+q^{-4})
\pmatrix{q^2 xy  & - q x^2 \cr
         q^2 y^2 & -   xy}  \xi \eta.                            \eqno(2.17)
$$

The operator $\pi_{12} D^2$ is a left-module morphism by construction.
One finds that the representation $\rho$ of (1.25) is given by
$\rho(f)(x^i) = f(q^2 x^i)$. Since $\Omega^3$ vanishes the Bianchi
identities are trivially satisfied.

The metric is a $\Omega^0$-linear map from
$\Omega^1 \otimes_{\Omega^0} \Omega^1$ into $\Omega^0$ which satisfies
(1.27).  It is straightforward to see that there can be no metric. One
can consistently impose the condition $g \sigma = - g$ but this
map resembles rather a symplectic form.  In the limit $q = 1$ there is a
metric, the ordinary euclidean metric, but the connection (2.1.2) is not
a metric connection. This can be seen from the absence of any symmetry
in the matrix on the right-hand side of (2.17).

\parskip 7pt plus 1pt
\parindent=0cm
%{\it Acknowledgment:}\
\vskip 1cm

\beginsection References

Chamseddine A.H., Felder G., Fr\"ohlich J. 1993, {\it Gravity in
Non-Commutative Geometry}, Commun. Math. Phys. {\bf 155} 205.

Connes A. 1986, {\it Non-Commutative Differential Geometry}, Publications
of the Inst. des Hautes Etudes Scientifique. {\bf 62} 257.

--- 1990, {\it G\'eom\'etrie noncommutative}, InterEditions, Paris.

Dubois-Violette M. 1988, {\it D\'erivations et calcul diff\'erentiel
non-commutatif}, C. R. Acad. Sci. Paris {\bf 307} S\'erie I 403.

Dubois-Violette M., Michor P. 1994a, {\it D\'erivations et calcul
diff\'erentiel non-commuta\-tif II}, C. R. Acad. Sci. Paris, (to appear).

Dubois-Violette M., Michor P. 1994b, {\it Connections in Central
Bimodules}, Orsay Preprint.

Koszul J.L. 1960, {\it Lectures on Fibre Bundles and Differential Geometry},
Tata Institute of Fundamental Research, Bombay.

Maltsiniotis G. 1993, {\it Le Langage des Espaces et des Groupes
Quantiques}, Commun. Math. Phys. {\bf 151} 275.

Manin Yu. I. 1989, {\it Multiparametric Quantum Deformations of the General
Linear Supergroup}, Commun. Math. Phys. {\bf 123} 163.

Mourad. J. 1994, {\it Linear Connections in Non-Commutative Geometry},
Univ. of Tour Preprint.

Pusz W., Woronowicz S.L. 1989, {\it Twisted Second Quantization}, Rep. on
Math. Phys. {\bf 27} 231.

Wess J., Zumino B. 1990, {\it Covariant Differential Calculus on the Quantum
Hyperplane} Nucl. Phys. B (Proc. Suppl.) {\bf 18B} 302.

Woronowicz S. L. 1987, {\it Twisted $SU(2)$ Group. An example of a
Non-Commutative Differential Calculus}, Publ. RIMS, Kyoto Univ. {\bf 23} 117.

\bye